\documentstyle[11pt,epic,eepic]{article}

\def\beq{\begin{equation}}
\def\eeq{\end{equation}}
\def\bea{\begin{eqnarray}}
\def\eea{\end{eqnarray}}

\def\nn{\nonumber}
\def\ox{\otimes}
\def\ld{\lambda}
\def\lm{\lambda}

\title
{\LARGE\bf The Oblique corrections \\from  The Diagonal ETC Interaction}
\bigskip
\bigskip
\vspace{1cm}
\author{ Tadashi YOSHIKAWA\thanks{e-mail:
yosikawa@theo.phys.sci.hiroshima-u.ac.jp}  \\
                             \\
\normalsize \em Department of Physics, Hiroshima University\\
\normalsize \em 1-3-1 Kagamiyama, Higashi-Hiroshima, 724\\
\normalsize \em Japan}

\date{}
\begin{document}
\setlength{\baselineskip}{24pt}
\maketitle
\begin{picture}(0,0)
\put(325,240){HUPD-9514}
\put(325,225){June, 1995}
\end{picture}
\vspace{-24pt}

\vspace{3cm}
\begin{abstract} \large
We study the effect of the diagonal extended
technicolor(ETC)
 gauge boson on the
oblique correction parameters.
It is shown that in the $T$ parameter is unacceptably large
when the $Zbb$ vertex correction and $S$ parameter are
consistent with the experiments in the ETC model.
\end{abstract}
\newpage
In the recent works \cite{wu}, it is shown that the diagonal
extended technicolor(ETC) interaction may solve the $Zbb$
problem, i.e., the discrepancy between the
experiment and the prediction of the Standard Model (SM) in
$Zbb$ vertex.
If the contribution of the
diagonal interaction to $Zbb$ vertex
is large enough to cancel the other corrections
for the $Zbb$ vertex, the discrepancy could be explained. However,
such large effect must contribute to the oblique
corrections because the effect comes from the breaking of the custodial
symmetry in the right handed ETC interaction. It is
necessary to break the custodial symmetry to generate the mass
difference between top and bottom quarks. Hence, the $T$
parameter must receive large contribution from the ETC
interactions. The diagram such as Fig.2\footnote{In
Ref.\cite{wu}, the
contribution from Fig.2(A) has been calculated but the contribution
from Fig.2(B) is not considered. }must contribute to the
oblique correction S,T and U \cite{pes}. In this letter, we study the
effect of the diagonal ETC interaction for the oblique
corrections in the case that the non-oblique correction of
the $Zbb$ vertex is consistent with
the experimental data in a realistic one-family model with the small
$S$ parameter\cite{apel}(the model without exact
custodial symmetry\cite{yosi}).

We study the model that the horizontal symmetry
$SU(N_{TC}+1)$ is broken into $SU(N_{TC})$. In the multiplet of
$SU(N_{TC}+1)$, the third generation of ordinary fermions and
the techni-fermions are contained.
The lagrangian for the diagonal ETC interaction in the
one-family technicolor model is
\bea
   {{\cal L}_{ETC (3-TC)}^D} =  g_{ETC}
                        X_{ETC}^\mu
                    \frac{1}{\sqrt{2 N_{TC}( N_{TC} +1 )}}
                &[& \xi_L^t ( \bar{Q}_L^i \gamma_\mu Q_L^i
                      - N_{TC}\bar{q}_L^i \gamma_\mu q_L^i )\nn \\
                &+&  \xi_R^t ( \bar{U}_R^i \gamma_\mu U_R^i
                      - N_{TC}\bar{t}_R^i \gamma_\mu t_R^i )\nn\\
                &+&  \xi_R^b ( \bar{D}_R^i \gamma_\mu D_R^i
                      - N_{TC}\bar{b}_R^i \gamma_\mu b_R^i )]\nn \\
                &+& \xi_L^\tau ( \bar{L}_L \gamma_\mu L_L
                      - N_{TC}\bar{l}_L \gamma_\mu l_L )\nn \\
                &+&  \xi_R^\nu ( \bar{N}_R \gamma_\mu N_R
                      - N_{TC}\bar{\nu}_R \gamma_\mu \nu_R )\nn\\
                &+&  \xi_R^\tau ( \bar{E}_R \gamma_\mu E_R
                    - N_{TC}\bar{\tau}_R \gamma_\mu \tau_R )]
\eea
where $Q_L^i = ( U^i,D^i )_L$, $U_R^i$ and $D_R^i$ represent
techniquarks,
$q_L^i = ( t^i,b^i )_L$, $t_R^i$ and $b_R^i$ represent
the third family of quarks and `` i ''
is the color index of QCD. $L_L = ( N,E )_L$, $E_R$
represent the technilepton, $l_L = ( \nu,\tau )_L$
and $\tau_R$ represent the third family of leptons. $g_{ETC}$
is a coupling of ETC interaction. $X_{ETC}$
is diagonal ETC gauge boson which mediates between the third
family of ordinary fermions and techni
fermions. $N_{TC}$ is the number of the technicolor.
$ \frac{1}{\sqrt{2 N_{TC}( N_{TC} +1 )}}$ is the
normalization factor of the generator of horizontal symmetry
$SU(N_{TC} + 1)$.
$\xi_L^{t(\tau)}$ is a coefficient of left handed
coupling and $\xi_R^{t(b,\tau)}$ is one of right handed
coupling. Since the left handed fermion belongs to
$SU(2)$ doublet, the couplings of up-type quark and
down-type quark
in the doublet are the same as each other.

The effective lagrangian for fig.1 is
\bea
{\cal L}_{int} =
            \frac{1}{2} \frac{g_{ETC}^2 }{q^2-M_{X}^2}
            \frac{1}{2 N_{TC}(N_{TC}+1)} &[& {\xi_L^t}
                     \bar{Q}_L^i \gamma^\mu Q_L^i
              + \xi_R^t \bar{U}_R^i
                                   \gamma^\mu U_R^i
              + \xi_R^b \bar{D}_R^i
                                \gamma^\mu D_R^i
                                  \nn \\
             &-&N_{TC}\xi_L^t \bar{q}_L^i \gamma_\mu
                                   q_L^i
            -N_{TC}\xi_R^t \bar{t}_R^i \gamma_\mu
                                   t_R^i
            -N_{TC}\xi_R^b \bar{b}_R^i \gamma_\mu
                                   b_R^i \nn \\
             &+& \xi_L^\tau \bar{L}_L \gamma^\mu L_L
              + \xi_R^\nu \bar{N}_R \gamma^\mu N_R
              + \xi_R^\tau \bar{E}_R \gamma^\mu E_R \\
            &-&N_{TC}\xi_L^\tau \bar{l}_L \gamma_\mu
                                   l_L
            -N_{TC}\xi_R^\nu \bar{\nu}_R \gamma_\mu
                                   \nu_R
            -N_{TC}\xi_R^\tau \bar{\tau}_R \gamma_\mu
                                   \tau_R \nn
                      ]^2,
\eea
where $M_X$ is the mass of ETC gauge boson.
Below the TC chiral symmetry breaking scale, the current of techniquarks
are replaced by the Noether
current\cite{gio,civ,kit,kit2,yosi2}
 in the effective chiral lagrangian with
$SU(2N_c)_L \ox SU(2N_c)_R \ox U(1)_V$ in techniquark sector
\cite{yosi, yosi2}.
Here, we separate the
right-handed current into $\tau^3$ and singlet components
of $SU(2)$,
\bea
 \xi_R^t \bar{U}_R \gamma^\mu U_R
              + \xi_R^b \bar{D}_R \gamma^\mu D_R = \frac{
 \xi_R^t + \xi_R^b }{2} \bar{Q}_R \gamma^\mu Q_R
+ \frac{ \xi_R^t - \xi_R^b }{2} \bar{Q}_R \tau^3 \gamma^\mu
                           Q_R
\eea
Explicitly, the right-handed currents of techniquark
are replaced by the following Noether current of the
effective lagrangian.
\bea
\sum_{i=1}^{3} \bar{Q}_R^i \gamma^\mu Q_R^i &\sim&
          - 3 \frac{M_{6\omega}}{G_{6\omega}}
                 [ \omega_6^\mu  - \frac{\sqrt{3}}{2 G_{6\omega}}
                    2 Y_{Lq} g^\prime B^\mu ]
                        \frac{Y_{Lq}}{\sqrt{3}}, \\
\sum_{i=1}^{3} \bar{Q}_R^i \tau^3 \gamma^\mu Q_R^i &\sim&
          3 F_6^2 \frac{1}{2} ( g W^\mu - g^\prime B^\mu )
                  \nn \\
 & & - 3 \frac{M_{V6}}{G_{6}}
                 [ \rho_6^\mu  - \frac{\sqrt{3}}{2 G_6}
                    ( g W^\mu + g^\prime B^\mu) ]
                                \frac{1}{\sqrt{3}}\nn
                                     \\
 & & - 3 \frac{M_{A6}}{\ld_{6}}
                 [ a_6^\mu  + \frac{\sqrt{3}}{2 \ld_6}
                    ( g W^\mu - g^\prime B^\mu) ]
                                \frac{1}{\sqrt{3}},
\eea
where, $\omega_\mu$ and $\rho_\mu$ are techni-omega meson
and techni-rho meson that is composed by techniquarks and
$M_{\omega6}$ and $M_{V6}$ are their masses.
 $a_\mu$ is a techni-axialvectormeson and
$M_{A6}$ is its mass. $G$ and $\ld$ are the couplings which
are related to the techni-vectormesons.
 The $F_6$ is the decay constant of technipion
in techniquark sector.
We can neglect the technilepton contribution to the oblique
corrections because the
coefficients of ETC coupling or decay constant $F_2$
in technilepon sector is much smaller
than that of techniquark in order to generate the
mass difference between techniquark and technilepton.
Besides this reason, in the
model with small $S$ parameter\cite{apel}\cite{yosi},
the decay constant
$F_2$ must be much smaller than that in the techniquark
sector to
satisfy the experimental bound of $T$ parameter,.

The main part of the contributions to oblique correction
from the diagonal ETC
interaction (Fig.2(A)) is
\bea
      \frac{9}{64} \frac{g_{ETC}^2}{p^2 - M_X^2}
         \frac{1}{N_{TC}(N_{TC}+1)}
          (\xi_R^t - \xi_R^b)^2
              F_\pi^4
          (gW_3 - g^\prime B)^2 .
\eea
The contribution from the techni(axial)vectormesons
is also given by,
\bea
      \frac{9}{32}\frac{g_{ETC}^2}{p^2 - M_X^2}
         \frac{1}{N_{TC}(N_{TC}+1)}
          (\xi_R^t - \xi_R^b)^2& &
                F_\pi^2
          (gW_3 - g^\prime B) \\
  \times &\{& [ \frac{M_V^2}{G^2} + \frac{M_V^2}{G}
                   \frac{1}{p^2 - M_V^2}
                   \frac{M_V^2}{G} ](gW_3 + g^\prime B) \nn
                               \\
         &-& [  \frac{M_A^2}{\lm^2} + \frac{M_A^2}{\lm}
                   \frac{1}{p^2 - M_A^2}
                   \frac{M_A^2}{\lm} ]
                            (gW_3 - g^\prime B)\}\nn \\
 + \frac{9}{64} \frac{g_{ETC}^2}{p^2 - M_X^2}
         \frac{1}{N_{TC}(N_{TC}+1)}
          (\xi_R^t - \xi_R^b)^2
   &\{& [ \frac{M_V^2}{G^2} + \frac{M_V^2}{G}
                   \frac{1}{p^2 - M_V^2}
                   \frac{M_V^2}{G} ](gW_3 + g^\prime B) \nn
                               \\
         &-& [  \frac{M_A^2}{\lm^2} + \frac{M_A^2}{\lm}
                   \frac{1}{p^2 - M_A^2}
                   \frac{M_A^2}{\lm} ]
                            (gW_3 - g^\prime B)\}^2.\nn
\eea


Using eq.(6) and eq.(7), we obtain the contributions to the oblique
parameters\cite{pes} from Fig.2(A)
\bea
S^{dETC} &=& - \frac{9}{2} \pi \frac{(\xi_R^t - \xi_R^b)^2}{N_{TC}(N_{TC}+1)}
             [\frac{g_{ETC}^2}{M_X^4} F_\pi^4
              + 2
\frac{g_{ETC}^2}{M_X^2}\frac{1}{\lm^2}F_\pi^2],
\label{s}
\\
\alpha T_{A}^{ETC} &=& \frac{9}{32}
             \frac{(\xi_R^t - \xi_R^b)^2}{N_{TC}(N_{TC}+1)}
             \frac{g_{ETC}^2}{M_X^2} F_\pi^4
                  \frac{g^2 + {g^\prime}^2}{M_Z^2}.
\eea
{}From this analysis, the contribution to
the $S$ parameter for the diagonal ETC gauge
interaction is negative\footnote{In this letter, we only
consider the contribution from techniquarks.
The $S^{ETC}$ of eq.(\ref{s})
has negative sign but the contribution is small compared
with that to the
$T$ parameter (See Fig. 5.).
However there may be the large contribution to $S$ from the other
fermions.}.

There is also another two loop contribution to the oblique correction
$T$ from the diagonal ETC interaction (Fig. 2(B)).
Below the ETC scale, the contribution is obtained from
the following four-fermi
lagrangian:
\bea
-  \frac{1}{4}\frac{g_{ETC}^2}{M_X^2}\frac{1}{N_{TC}+1}
       &[& {\xi_L^t}^2 ( \bar{Q}_L^i \gamma_\mu Q_L^i )^2 +
\nn \\
       &+& \frac{
         (\xi_R^t + \xi_R^b)^2 }{4} (\bar{Q}_R \gamma^\mu Q_R)^2
        +  \frac{ (\xi_R^t - \xi_R^b)^2 }{4}
         (\bar{Q}_R \tau^3 \gamma^\mu Q_R )^2.
\eea
After Fierz transformation, the lagrangian becomes to
\bea
- \frac{1}{4}\frac{g_{ETC}^2}{M_X^2}\frac{1}{N_{TC}+1}
       &[& \frac{{\xi_L^t}^2}{2} \sum_{A=0}^{3}
          (\bar{Q}_L^i \gamma_\mu \tau^A Q_L^i )^2 +
                        \nn \\
       &+& \frac{
         (\xi_R^t + \xi_R^b)^2 }{8} \sum_{A=0}^{3}
               (\bar{Q}_R \gamma^\mu \tau^A Q_R)^2 \\
       &+&  \frac{ (\xi_R^t - \xi_R^b)^2 }{8} \{
         (\bar{Q}_R \gamma^\mu Q_R )^2 + (\bar{Q}_R
                        \gamma^\mu \tau^3 Q_R )^2
        - \sum_{a=1}^{2}( \bar{Q}_R \gamma_\mu \tau^a Q_L^i
)^2\} ], \nn
\eea
where, $\tau^a (a=1,2,3)$ is the Pauli matrix and $\tau^0$
is a unit matrix.
Note that the sign in the third term different with the
other terms. We replace the currents of technifermion
by the Noether current. Then, the contribution to $T$ from
Fig.2(B) is given by
\bea
&-& \frac{3}{32}\frac{g_{ETC}^2}{M_X^2}\frac{F_\pi^4}{N_{TC}+1}
  [ \xi_L^2 + \frac{ (\xi_R^t + \xi_R^b)^2 }{4} +
               \frac{(\xi_R^t - \xi_R^b)^2 }{4}](gW_3 -
g^\prime B)^2 \nn \\
&-& \frac{3}{32}\frac{g_{ETC}^2}{M_X^2}\frac{F_\pi^4}{N_{TC}+1}
  [ \xi_L^2 + \frac{ (\xi_R^t + \xi_R^b)^2 }{4} -
               \frac{(\xi_R^t - \xi_R^b)^2 }{4}]\sum_{a=1}^2
( gW^a )^2
\eea
Hence, only the terms of a factor of $(\xi_R^t - \xi_R^b)^2$
only contribute
to $T$ from Fig.2(B).
The contribution to $T$ parameter is
\bea
\alpha T_{B}^{ETC} &=& \frac{3}{32}
             \frac{(\xi_R^t - \xi_R^b)^2}{N_{TC}(N_{TC}+1)}
             \frac{g_{ETC}^2}{M_X^2} F_\pi^4
                  \frac{g^2 + {g^\prime}^2}{M_Z^2}.
\eea
The total contribution to $T$ from the diagonal ETC interaction is
\bea
T^{ETC} = T_{A}^{ETC} + T_{B}^{ETC}.
\label{t}
\eea

While, the non-oblique corrections for $Zbb$ vertex
\cite{wu}
are given by
\bea
\delta g_{L}^{ETC} = \delta g_{LS}^{ETC} + \delta
g_{LD}^{ETC},
\label{gs+gd}
\eea
where, the contribution from
the side-way ETC gauge interaction of Fig.3 is
\bea
\delta g_{LS}^{ETC} &=& \frac{1}{8}{\xi_L^t}^2
        \frac{g_{ETC}^2}{M_{ETC}^2} F_\pi^2
         \sqrt{g^2 + {g^\prime}^2},
\eea
and the contribution from the diagonal ETC interaction of
Fig.4 is
\bea
\delta g_{LD}^{ETC} &=& - \frac{3}{8}\xi_L^t
           (\xi_R^t - \xi_R^b) \frac{g_{ETC}^2}{M_{ETC}^2}
         \frac{1}{N_{TC}+1} F_\pi^2
         \sqrt{g^2 + {g^\prime}^2}.
\eea
If the effect of the ETC gauge interaction, i.e.,
eq.(\ref{gs+gd}) explain
the difference
between the experimental data of $R_b$ and the prediction of
SM,
the parameter $\xi_R^t - \xi_R^b$ must be larger than
$\xi_L(N_{TC}+1) /3$
and small $M_X/g_{ETC}$ is favored.
Since $S$ parameter is proportional to $N_{TC}$, the small
$N_{TC}$ is favored to be consistent with the experimental
constraint for $S$.
Therefore we choose  $N_{TC}=2$.
The parameter $\xi_L^t$ is taken to be unity for simplicity.
Comparing
the mass of
between top quark and bottom quark, $\xi_R^t$ is much larger
than $\xi_R^b$.
Hence, we treat $\xi_R^t$ as the
parameter which show the breaking of custodial symmetry.
In the model with small $S$
parameter\cite{apel}
( the model without exact custodial symmetry\cite{yosi} ),
$F_{\pi} \sim \sqrt{250^2/3}
\sim 144GeV$. In eq.(\ref{s}), we put $\ld^2 = 106$ (See ref.\cite{yosi}).
Here, we define a ratio of the ETC correction to $R_b$
\bea
\frac{\delta R_b^{ETC}}{R_b} \sim (1-R_b)
               \frac{2 g_L \delta g_L^{ETC}}{g_L^2 + g_R^2}.\nn
\eea
In Fig.5, the ratio presented as the functions of
$M_{X}/g_{ETC}$
for several values of $\xi_R^t$.
Because the $\xi_R^t$ must be larger than $\xi_L(N_{TC}+1)/3
= 1$, we choose the following values for
$\xi_R^t - \xi_R^b \sim \xi_R^t$. (a)1.2, (b)1.5, (c)2 and
(d)2.5.
If the contribution to $R_b$ from the ETC model explains
the experimental data in 1 $\sigma$ level, the $\delta
R_b^{ETC}/R_b$ must larger than about 0.012. Then, in Fig.5,
it is shown that
the mass of ETC gauge boson $M_X/g_{ETC}$ must be smaller than
about 700 GeV in case (b), 900 GeV in (c), 1100 GeV in (d).

In Fig.6 and Fig.7, we plot the behavior of the
contribution to oblique correction from diagonal ETC
interaction (eq.(\ref{s}) and eq.(\ref{t})),
by choosing the same values for $\xi_R^t$ as those in Fig.5.
For the values of $M_x/g_{ETC}$ which satisfy
the experimental constraint of $R_b$, the contribution to $S$ from
ETC negligible compared with that from TC (The tipical TC
cobtribution to $S$ is $0.1 N_{TC}$ from a one doublet
technifermion.).
$T$ receive large
value.
In Fig.7, it is shown that the value of $T$ must be larger than
about 0.9 in the cases (b),(c) and (d) for 1
$\sigma$ level of experiment of $R_b$.
This value contradict with  the experimental bound of $T$
( $T_{exp} < 0.5$).
In the model with  small $S$\cite{apel}, the situation
is worse because $T$ parameter already receives the
contribution from the custodial symmetry breaking in
technilepton sector.
Hence, it is not favored that
the $T$ receives the additional contribution from ETC interaction.
It is difficult that the
discrepancy between the SM and the experiment for the $R_b$
is explained by the
contribution of
the diagonal ETC gauge
interaction, because the contribution to $T$ parameter
contradicts with the experimental bound.

The contribution to the vertex correction of $Zbb$
from the diagonal ETC gauge interaction
become large with positive sign when the $\xi_R^t - \xi_R^b$
is larger than $\xi_L^t (N_{TC}+1)/3 $. However, because the such
large $\xi_R^t - \xi_R^b$ breaks the custodial
symmetry significantly,
$T$ must receive the contribution from the diagonal ETC
interaction.
It is difficult that
the $\xi_R^t g_{ETC}/ M_X$ becomes large enough to
explain  the discrepancy for $R_b$ ,
unless the other mechanism suppress the $T$ parameter
in this model.

{\Large{\bf Acknowledgement}}\\
We would like to thank Dr.T.Morozumi for
discussion and reading the manuscript.

\newpage
\begin{center}
{\Large{\bf Figure Captions}}
\end{center}
\begin{itemize}
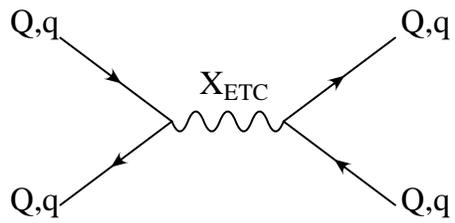
\item {\bf Fig. 1 }: The Feynman diagram for the
diagonal ETC gauge interaction.

\item {\bf Fig. 2 (A),(B)}: The Feynman diagrams for the
contribution to the oblique correction according to diagonal
ETC gauge interaction.

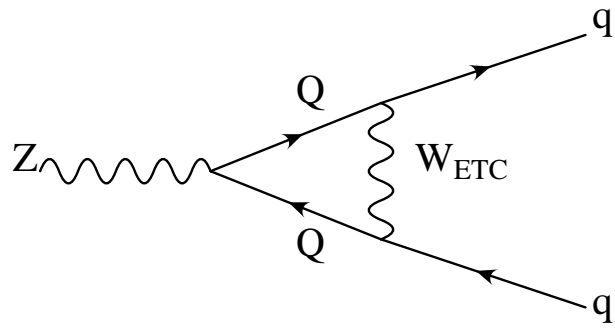
\item {\bf Fig. 3 }: The Feynman diagram for the
contribution to the vertex correction according to diagonal
ETC gauge interaction.

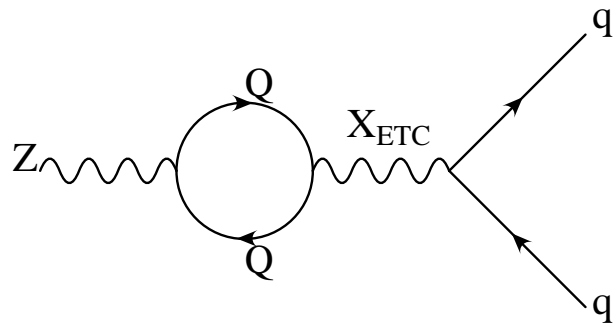
\item {\bf Fig. 4 }: The Feynman diagram for the
contribution to the oblique correction according to sideway
ETC gauge interaction.




\item {\bf Fig. 5,6,7}: $\frac{\delta R_b^{ETC}}{R_b}$, $S^{ETC}$
and $T^{ETC}$ as a function of $M_{X}/g_{ETC}$ for
following values for $\xi_R^t - \xi_R^b$. (a)
1.2 with a dashed thinline, (b) 1.5 with a thinline, (c) 2
with a thickline and (d) 2.5 with a dashed thickline.
\end{itemize}
\newpage

\begin{figure}
\setlength{\unitlength}{0.240900pt}
\begin{picture}(1500,900)(0,0)
\tenrm
\thicklines \path(220,113)(240,113)
\thicklines \path(1436,113)(1416,113)
\put(198,113){\makebox(0,0)[r]{0}}
\thicklines \path(220,240)(240,240)
\thicklines \path(1436,240)(1416,240)
\thicklines \path(220,368)(240,368)
\thicklines \path(1436,368)(1416,368)
\put(198,368){\makebox(0,0)[r]{0.01}}
\thicklines \path(220,495)(240,495)
\thicklines \path(1436,495)(1416,495)
\thicklines \path(220,622)(240,622)
\thicklines \path(1436,622)(1416,622)
\put(198,622){\makebox(0,0)[r]{0.02}}
\thicklines \path(220,750)(240,750)
\thicklines \path(1436,750)(1416,750)
\thicklines \path(220,877)(240,877)
\thicklines \path(1436,877)(1416,877)
\put(198,877){\makebox(0,0)[r]{0.03}}
\thicklines \path(301,113)(301,133)
\thicklines \path(301,877)(301,857)
\put(301,68){\makebox(0,0){600}}
\thicklines \path(463,113)(463,133)
\thicklines \path(463,877)(463,857)
\put(463,68){\makebox(0,0){800}}
\thicklines \path(625,113)(625,133)
\thicklines \path(625,877)(625,857)
\put(625,68){\makebox(0,0){1000}}
\thicklines \path(787,113)(787,133)
\thicklines \path(787,877)(787,857)
\put(787,68){\makebox(0,0){1200}}
\thicklines \path(950,113)(950,133)
\thicklines \path(950,877)(950,857)
\put(950,68){\makebox(0,0){1400}}
\thicklines \path(1112,113)(1112,133)
\thicklines \path(1112,877)(1112,857)
\put(1112,68){\makebox(0,0){1600}}
\thicklines \path(1274,113)(1274,133)
\thicklines \path(1274,877)(1274,857)
\put(1274,68){\makebox(0,0){1800}}
\thicklines \path(1436,113)(1436,133)
\thicklines \path(1436,877)(1436,857)
\put(1436,68){\makebox(0,0){2000}}
\thicklines \path(220,113)(1436,113)(1436,877)(220,877)(220,113)
\put(45,495){\makebox(0,0)[l]{\shortstack{$ \frac{\delta R_b^{ETC}}{R_b} $}}}
\put(828,23){\makebox(0,0){$\frac{M_{X}}{g_{ETC}}$ \ \ \  [GeV]}}
\thinlines  \path(220,301)(220,301)(271,262)(321,234)
(372,213)(423,197)(473,184)(524,175)(575,167)(625,160)
(676,155)(727,150)(777,146)(828,143)(879,140)(929,138)
(980,136)(1031,134)(1081,132)(1132,131)(1183,130)(1233,128)
(1284,127)(1335,126)(1385,126)(1436,125)
\thinlines  \dashline[-10]{25}(220,584)(220,584)(271,485)
(321,414)(372,362)(423,322)(473,291)(524,267)(575,247)
(625,231)(676,217)(727,206)(777,196)(828,188)(879,181)
(929,175)(980,170)(1031,165)(1081,161)(1132,158)(1183,154)
(1233,151)(1284,149)(1335,146)(1385,144)(1436,142)
\Thicklines  \path(266,877)(271,857)(321,716)(372,611)
(423,532)(473,470)(524,421)(575,381)(625,348)(676,322)
(727,299)(777,280)(828,264)(879,250)(929,238)(980,227)
(1031,218)(1081,209)(1132,202)(1183,196)(1233,190)(1284,185)
(1335,180)(1385,176)(1436,172)
\Thicklines  \dashline[-10]{25}(367,877)(372,860)(423,741)
(473,648)(524,574)(575,515)(625,466)(676,426)(727,392)
(777,363)(828,339)(879,318)(929,300)(980,284)(1031,270)
(1081,258)(1132,247)(1183,237)(1233,228)(1284,221)
(1335,213)(1385,207)(1436,201)
\end{picture}

\vspace{0.7cm}
\begin{center}
{\bf Fig.5}
\end{center}
\end{figure}

\begin{figure}
\setlength{\unitlength}{0.240900pt}
\begin{picture}(1500,900)(0,0)
\tenrm
\thicklines \path(220,113)(240,113)
\thicklines \path(1436,113)(1416,113)
\put(198,113){\makebox(0,0)[r]{-0.1}}
\thicklines \path(220,189)(240,189)
\thicklines \path(1436,189)(1416,189)
\thicklines \path(220,266)(240,266)
\thicklines \path(1436,266)(1416,266)
\put(198,266){\makebox(0,0)[r]{-0.08}}
\thicklines \path(220,342)(240,342)
\thicklines \path(1436,342)(1416,342)
\thicklines \path(220,419)(240,419)
\thicklines \path(1436,419)(1416,419)
\put(198,419){\makebox(0,0)[r]{-0.06}}
\thicklines \path(220,495)(240,495)
\thicklines \path(1436,495)(1416,495)
\thicklines \path(220,571)(240,571)
\thicklines \path(1436,571)(1416,571)
\put(198,571){\makebox(0,0)[r]{-0.04}}
\thicklines \path(220,648)(240,648)
\thicklines \path(1436,648)(1416,648)
\thicklines \path(220,724)(240,724)
\thicklines \path(1436,724)(1416,724)
\put(198,724){\makebox(0,0)[r]{-0.02}}
\thicklines \path(220,801)(240,801)
\thicklines \path(1436,801)(1416,801)
\thicklines \path(220,877)(240,877)
\thicklines \path(1436,877)(1416,877)
\put(198,877){\makebox(0,0)[r]{0}}
\thicklines \path(301,113)(301,133)
\thicklines \path(301,877)(301,857)
\put(301,68){\makebox(0,0){600}}
\thicklines \path(463,113)(463,133)
\thicklines \path(463,877)(463,857)
\put(463,68){\makebox(0,0){800}}
\thicklines \path(625,113)(625,133)
\thicklines \path(625,877)(625,857)
\put(625,68){\makebox(0,0){1000}}
\thicklines \path(787,113)(787,133)
\thicklines \path(787,877)(787,857)
\put(787,68){\makebox(0,0){1200}}
\thicklines \path(950,113)(950,133)
\thicklines \path(950,877)(950,857)
\put(950,68){\makebox(0,0){1400}}
\thicklines \path(1112,113)(1112,133)
\thicklines \path(1112,877)(1112,857)
\put(1112,68){\makebox(0,0){1600}}
\thicklines \path(1274,113)(1274,133)
\thicklines \path(1274,877)(1274,857)
\put(1274,68){\makebox(0,0){1800}}
\thicklines \path(1436,113)(1436,133)
\thicklines \path(1436,877)(1436,857)
\put(1436,68){\makebox(0,0){2000}}
\thicklines \path(220,113)(1436,113)(1436,877)(220,877)(220,113)
\put(45,495){\makebox(0,0)[l]{\shortstack{$ S^{ETC} $}}}
\put(828,23){\makebox(0,0){$\frac{M_{X}}{g_{ETC}}$ \ \ \  [GeV]}}
\thinlines \path(220,698)(220,698)(245,737)(271,765)
(296,787)(321,804)(347,817)(372,827)(397,835)(423,842)
(448,847)(473,851)(499,855)(524,858)(575,862)(625,866)
(676,868)(727,870)(777,871)(828,872)(879,873)(929,874)
(980,874)(1031,875)(1081,875)(1132,875)(1183,876)(1233,876)
(1284,876)(1335,876)(1385,876)(1436,876)
\thinlines  \dashline[-10]{25}(220,598)(220,598)(245,658)
(271,703)(296,737)(321,763)(347,783)(372,799)(397,811)
(423,822)(448,830)(473,837)(499,842)(524,847)(575,854)
(625,859)(676,863)(727,866)(777,868)(828,870)(879,871)
(929,872)(980,873)(1031,873)(1081,874)(1132,874)(1183,875)
(1233,875)(1284,875)(1335,876)(1385,876)(1436,876)
\Thicklines \path(220,381)(220,381)(245,488)(271,567)
(296,627)(321,674)(347,710)(372,738)(397,761)(423,779)
(448,793)(473,806)(499,816)(524,824)(575,837)(625,846)
(676,852)(727,857)(777,861)(828,864)(879,866)(929,868)
(980,870)(1031,871)(1081,872)(1132,872)(1183,873)(1233,874)
(1284,874)(1335,874)(1385,875)(1436,875)
\Thicklines  \dashline[-10]{25}(222,113)(245,268)
(271,393)(296,487)(321,559)(347,615)(372,660)(397,695)
(423,723)(448,747)(473,765)(499,781)(524,794)(575,814)
(625,828)(676,839)(727,846)(777,852)(828,857)(879,860)
(929,863)(980,865)(1031,867)(1081,869)(1132,870)(1183,871)
(1233,872)(1284,872)(1335,873)(1385,873)(1436,874)
\end{picture}

\vspace{0.7cm}
\begin{center}
{\bf Fig.6}
\end{center}
\end{figure}

\begin{figure}
\setlength{\unitlength}{0.240900pt}
\begin{picture}(1500,900)(0,0)
\tenrm
\thicklines \path(220,113)(240,113)
\thicklines \path(1436,113)(1416,113)
\put(198,113){\makebox(0,0)[r]{0}}
\thicklines \path(220,189)(240,189)
\thicklines \path(1436,189)(1416,189)
\put(198,189){\makebox(0,0)[r]{0.2}}
\thicklines \path(220,266)(240,266)
\thicklines \path(1436,266)(1416,266)
\put(198,266){\makebox(0,0)[r]{0.4}}
\thicklines \path(220,342)(240,342)
\thicklines \path(1436,342)(1416,342)
\put(198,342){\makebox(0,0)[r]{0.6}}
\thicklines \path(220,419)(240,419)
\thicklines \path(1436,419)(1416,419)
\put(198,419){\makebox(0,0)[r]{0.8}}
\thicklines \path(220,495)(240,495)
\thicklines \path(1436,495)(1416,495)
\put(198,495){\makebox(0,0)[r]{1}}
\thicklines \path(220,571)(240,571)
\thicklines \path(1436,571)(1416,571)
\put(198,571){\makebox(0,0)[r]{1.2}}
\thicklines \path(220,648)(240,648)
\thicklines \path(1436,648)(1416,648)
\put(198,648){\makebox(0,0)[r]{1.4}}
\thicklines \path(220,724)(240,724)
\thicklines \path(1436,724)(1416,724)
\put(198,724){\makebox(0,0)[r]{1.6}}
\thicklines \path(220,801)(240,801)
\thicklines \path(1436,801)(1416,801)
\put(198,801){\makebox(0,0)[r]{1.8}}
\thicklines \path(220,877)(240,877)
\thicklines \path(1436,877)(1416,877)
\put(198,877){\makebox(0,0)[r]{2}}
\thicklines \path(301,113)(301,133)
\thicklines \path(301,877)(301,857)
\put(301,68){\makebox(0,0){600}}
\thicklines \path(463,113)(463,133)
\thicklines \path(463,877)(463,857)
\put(463,68){\makebox(0,0){800}}
\thicklines \path(625,113)(625,133)
\thicklines \path(625,877)(625,857)
\put(625,68){\makebox(0,0){1000}}
\thicklines \path(787,113)(787,133)
\thicklines \path(787,877)(787,857)
\put(787,68){\makebox(0,0){1200}}
\thicklines \path(950,113)(950,133)
\thicklines \path(950,877)(950,857)
\put(950,68){\makebox(0,0){1400}}
\thicklines \path(1112,113)(1112,133)
\thicklines \path(1112,877)(1112,857)
\put(1112,68){\makebox(0,0){1600}}
\thicklines \path(1274,113)(1274,133)
\thicklines \path(1274,877)(1274,857)
\put(1274,68){\makebox(0,0){1800}}
\thicklines \path(1436,113)(1436,133)
\thicklines \path(1436,877)(1436,857)
\put(1436,68){\makebox(0,0){2000}}
\thicklines \path(220,113)(1436,113)(1436,877)(220,877)(220,113)
\put(45,495){\makebox(0,0)[l]{\shortstack{$ T^{ETC} $}}}
\put(828,23){\makebox(0,0){$\frac{M_{X}}{g_{ETC}}$ \ \ \  [GeV]}}
\thinlines \path(220,508)(220,508)(271,425)(321,366)
(372,322)(423,289)(473,263)(524,242)(575,225)(625,212)
(676,201)(727,191)(777,183)(828,176)(879,170)(929,165)
(980,161)(1031,157)(1081,153)(1132,150)(1183,148)(1233,145)
(1284,143)(1335,141)(1385,139)(1436,138)
\thinlines  \dashline[-10]{25}(220,731)(220,731)(271,601)
(321,508)(372,440)(423,388)(473,347)(524,315)(575,289)
(625,267)(676,250)(727,235)(777,223)(828,212)(879,203)
(929,195)(980,188)(1031,182)(1081,176)(1132,171)(1183,167)
(1233,163)(1284,160)(1335,157)(1385,154)(1436,152)
\Thicklines \path(303,877)(321,816)(372,694)(423,601)
(473,529)(524,472)(575,425)(625,388)(676,356)(727,330)
(777,308)(828,289)(879,272)(929,258)(980,246)(1031,235)
(1081,225)(1132,217)(1183,209)(1233,203)(1284,197)
(1335,191)(1385,186)(1436,182)
\Thicklines  \dashline[-10]{25}(422,877)(423,876)(473,763)
(524,673)(575,601)(625,542)(676,493)(727,452)(777,417)
(828,388)(879,362)(929,340)(980,321)(1031,304)(1081,289)
(1132,275)(1183,264)(1233,253)(1284,244)(1335,235)
(1385,227)(1436,220)
\end{picture}

\vspace{0.7cm}
\begin{center}
{\bf Fig.7}
\end{center}
\end{figure}


\begin{thebibliography}{99}
\bibitem{wu}
   G-H. Wu, {\it Phys. Rev. Lett.} {\bf 74} (1995) 4137,\\
   K. Hagiwara and N. Kitazawa, hep-ph/9504332.
\bibitem{pes}
   B. Holdom and J. Terning, {\it Phys. Lett.} {\bf B247}
                                           (1990) 88 ;\\
   M. Golden and L. Randall, {\it Nucl. Phys.} {\bf B361}
                                           (1991) 3 ;\\
   M. E. Peskin and T. Takeuchi, {\it Phys. Rev. Lett.} {\bf
                                   65} (1990) 964 ;
                                {\it Phys. Rev.} {\bf D46}
(1992) 381.
\bibitem{apel}
  T. Appelquist and J. Terning, {\it Phys. Lett.} {\bf B315} (1993) 139.
\bibitem{yosi}
   T. Yoshikawa, H. Takata and T. Morozumi, {\it Prog.
Theor. Phys. }{\bf 92} (1994) 353.
\bibitem{gio}
   H. Georgi, {\it Weak Interactions and Modern Particle
Theory}\\(Benjamin-Cummings,Menlo Park,1984),p77.
\bibitem{civ}
   R. S. Chivukula, E. B. Selipsky and E. H. Simmons, {\it
              Phys. Rev. Lett. }{\bf 69} (1992) 575 .
\bibitem{kit}
   R. S. Chivukula, E. Gates, E. H. Simmons and J. Terning,
{\it Phys. Lett. }{\bf B311} (1993) 157 ;\\
   N. Evans {\it Phys. Lett.} {\bf B331} (1994) 378 ;\\
   R. S. Chivukula, E. H. Simmons and J. Terning,
{\it Phys. Lett.} {\bf B331} (1994) 383.
\bibitem{kit2}
   N. Kitazawa, {\it Phys.  Lett. }{\bf B313} (1993) 395.
\bibitem{yosi2}
   T. Yoshikawa, hiroshima preprint HUPD-9415 hep-ph/9411280.
%
\end{thebibliography}
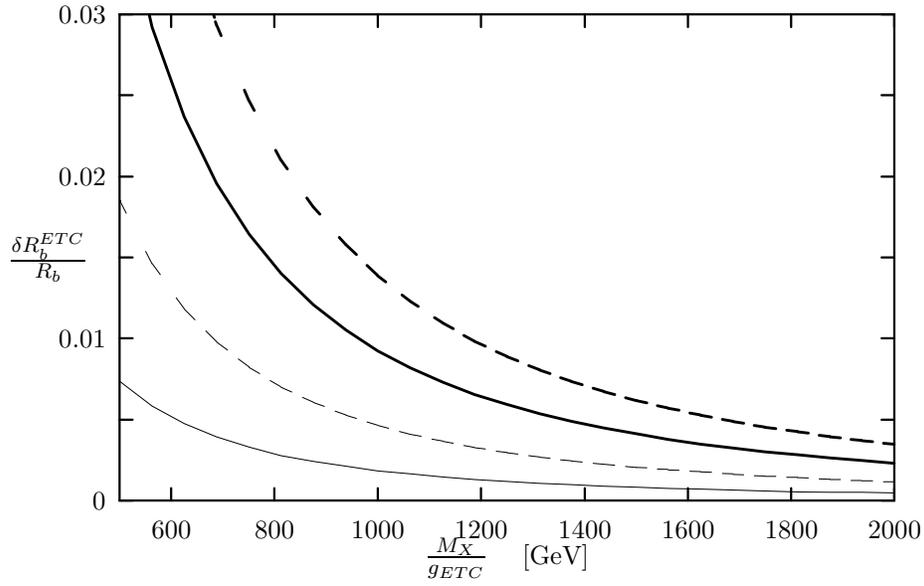
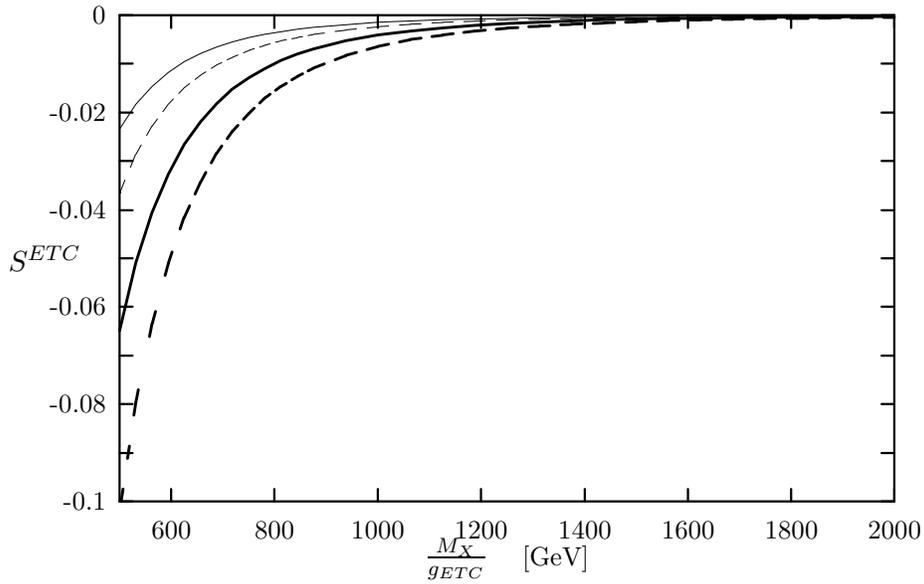
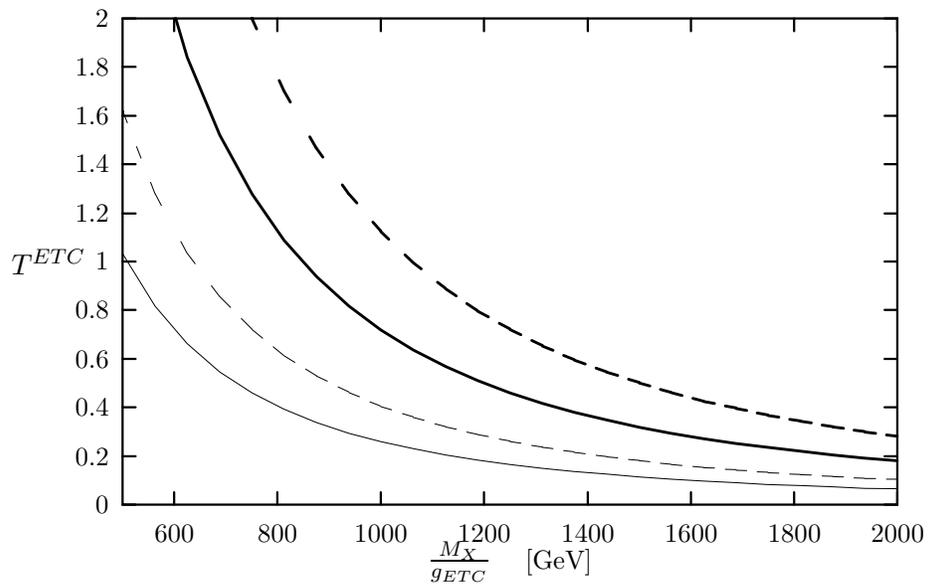
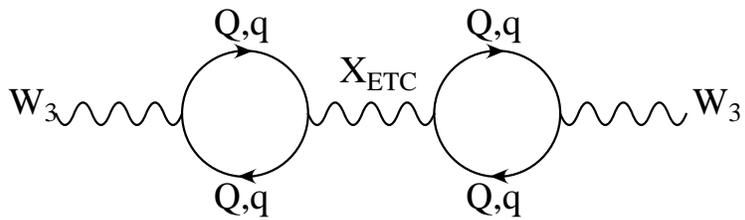
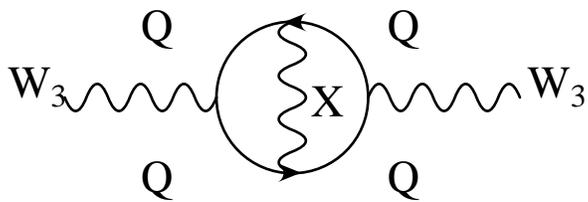
\end{document}